# Living with Neighbors. V. Better-aligned Spiral+Spiral Galaxy Pairs Show Stronger Star Formation

Woong-Bae G. Zee[1,2], Jun-Sung Moon[3,4], Sanjaya Paudel[1,2], and Suk-Jin Yoon[1,2]
[1] Department of Astronomy, Yonsei University, Seoul, 03722, Republic of Korea; sjyoon0691@yonsei.ac.kr
[2] Center for Galaxy Evolution Research, Yonsei University, Seoul, 03722, Republic of Korea
[3] Astronomy Program, Department of Physics and Astronomy, Seoul National University, Seoul, 08826, Republic of Korea
[4] Research Institute of Basic Sciences, Seoul National University, Seoul, 08826, Republic of Korea
*Received 2023 November 10; revised 2024 January 16; accepted 2024 January 17; published 2024 March 7*

## Abstract

Enhanced star formation (SF) with star-forming neighboring galaxies bolsters the hydrodynamical contributions during paired interactions. Although the relative spin orientation between interacting galaxies can influence this effect, it has not been comprehensively explored. In this study, utilizing a meticulously curated sample of nearby ($0.02 < z < 0.06$) spiral+spiral pairs and an isolated control sample from Sloan Digital Sky Survey Data Release 7, coupled with Galaxy Zoo 2, and an approach to estimate spin–spin alignment (SSA), we systematically compare the impact of relative orientation, $\cos \xi$, on interaction-induced SF. We reveal that SSA plays a pivotal role along with the conventionally recognized parameters of projected separation and SF of neighboring galaxies. Our results show an increase in SF augmentation as configurations transition from perpendicular ($\cos \xi \sim 0$) to well aligned ($\cos \xi > 0.7$). This enhancement is especially pronounced when neighboring galaxies have higher SF, suggesting that hydrodynamical processes drive the effects of relative orientation. The heightened hydrodynamical friction, attributed to increased ram pressure and the propensity for prograde orbits in well-aligned pairs, is consistent with our findings.

*Unified Astronomy Thesaurus concepts:* Galaxy evolution (594); Galaxy interactions (600); Galaxy pairs (610); Star formation (1569)

## 1. Introduction

The interactions between galaxies exert a significant influence on their evolution. Specifically, numerous observational studies have reported that galaxy–galaxy interactions can enhance the star formation rate (SFR; e.g., Yee & Ellingson 1995; Kennicutt et al. 1998; Ellison et al. 2008; Scudder et al. 2012; Patton et al. 2013; Cao et al. 2016; Silva et al. 2018; Pearson et al. 2019; Osborne et al. 2020; Shah et al. 2022). For example, using the Calar Alto Legacy Integral Field Area Survey (CALIFA), Barrera-Ballesteros et al. (2015) detected moderate star formation (SF) enhancement within interacting galaxies, typically localized in the central region. Steffen et al. (2021) also observed that the specific SFR (sSFR) in the central regions of interacting galaxies is enhanced by ∼0.3 dex, whereas a weak suppression is evident in their outer regions. This enhancement in sSFR is more pronounced in pairs with nearly equal masses and closer projected separations. Interaction-induced SF is commonly observed in merging galaxies (e.g., Boquien et al. 2009, 2010; Cortijo-Ferrero et al. 2017a, 2017b, 2017c; Thorp et al. 2019). Ellison et al. (2013) conducted an extensive study of ∼11,000 interacting/merging galaxies from the Sloan Digital Sky Survey (SDSS) and revealed that the distribution of interaction-induced SF enhancement is contingent on the evolutionary timescale of the galaxies. The pre-coalescence phase of interactions profoundly influences central SF, while the SFR expands in a broader region during the ultimate merging phase.

The phenomenon of interaction-induced SF activity in galaxy pairs has attracted increasing interest (e.g., Barton et al. 2000; Lambas et al. 2003; Woods & Geller 2007; Michel-Dansac et al. 2008; Woods et al. 2010; Patton et al. 2011; Shah et al. 2022). Barton et al. (2000) and Lambas et al. (2003), for example, provided statistical evidence of SF enhancement in galaxy pairs, discovering that increased SF becomes more pronounced as the projected distance between paired galaxies decreases. Michel-Dansac et al. (2008) revealed that less massive members in minor pairs (i.e., two galaxies with a high mass ratio) generally exhibit enriched metallicity, while major pair galaxies have a lower metallicity relative to isolated galaxies. They interpreted this finding as a result of the fact that metal-rich starburst events are often triggered by a more massive companion, and an inflow of low-metallicity fresh gas can be incited by a companion of similar or lesser mass. Similarly, Woods et al. (2010) associated the presence of starburst with major pairs, specifically within very close projected separation ($\Delta R < 30\ h^{-1}$ kpc). Recently, Shah et al. (2022) quantified the SF enhancement in ∼2400 major pairs at $0.5 < z < 3$ across the extensive COSMOS field, establishing that SF enhancement is most pronounced within the closest projected separation ($\Delta R < 25\ h^{-1}$ kpc). Furthermore, they determined that the level of SFR enhancement is dependent on redshift, in that SFR enhancement is ∼1.5 times higher at $0.5 < z < 1$ than at $1 < z < 3$ in the closest pairs, thereby indicating that the degree of interaction-induced SF significantly evolves with the age of the Universe.

The role of gas content during interactions remains an area of ongoing theoretical studies (e.g., Renaud et al. 2014; Moreno et al. 2015, 2019; Hani et al. 2020; Patton et al. 2020; Brown et al. 2023). Numerous simulations, incorporating a variety of merger scenarios with different mass ratios and gas fractions, have consistently shown that major mergers and gas-rich interactions notably amplify SF activity. The enhancement is primarily driven by the intense inflow of cold, dense gas, which fuels central SF activity (Bustamante et al. 2018). Additionally, Rupke et al. (2010) and Torrey et al. (2012) reproduced the dilution of metallicity and the flattening of radial gradients due to gas transfer during merger events. These simulations suggest that when gas-rich encounters occur, hydrodynamical friction, potentially induced by ram pressure, can cause material to lose its angular momentum, leading to increased gas inflow. This scenario has been established in several theoretical studies (Renaud et al. 2015, 2019; Capelo & Dotti 2017). Moster et al. (2011) and Gabor & Davé (2015) have demonstrated that a hot-gas halo can also potentially mitigate interaction-induced SF by inhibiting the influx of additional cold gas and preventing shock formation (Hwang & Park 2015; Correa et al. 2018a). Conversely, Karman et al. (2015) proposed that hot-gas halos initially enhance SF activity in the early stage of mergers by replenishing cold gas after the cooling of hot-gas reservoirs.

A growing body of observational research has focused on unraveling the role of hydrodynamical effects in interaction-induced SF activity in galaxy pairs (e.g., Woods & Geller 2007; Park & Choi 2009; Xu et al. 2010; Patton et al. 2011; Cao et al. 2016; Domingue et al. 2016; Zuo et al. 2018; He et al. 2022). For instance, Woods & Geller (2007) demonstrated a more prominent starburst occurs when both primary and secondary galaxies are blue, while red galaxies exhibit no enhancement in SF. Park & Choi (2009) further identified that late-type neighbors can enhance SF activity with remarkable efficiency, whereas early-type neighbors tend to suppress it, particularly in close pairs with a projected separation of galaxies within the virial radius. Xu et al. (2010) identified 27 close major pairs through the Two Micron All Sky Survey and SDSS Data Release 3, classifying them into spiral+spiral (S+S) and spiral+elliptical (S+E) pairs. Their findings revealed that the more massive ($M_*/M_\odot > 10^{11}$) spirals in S+S pairs exhibit, on average, ∼3 times higher sSFRs than isolated galaxies, while spirals in S+E pairs do not.

Building on this, Cao et al. (2016) observed that star-forming galaxies (SFGs) in S+S pairs exhibit a significantly augmented sSFR and SF efficiency (SFE), whereas there is neither an enhancement in sSFR nor an increase in SFE for S+E pairs. Moreover, SFGs in S+E pairs were found to possess marginally lower gas masses in comparison to their counterparts in both S+S pairs and isolated samples. Domingue et al. (2016) detected an overabundance of interstellar radiation fields and SF within spirals in S+S pairs, a phenomenon absent in S+E pairs. Zuo et al. (2018) examined 88 close major pairs through observations by Herschel and, as a supplement, H I data from the Green Bank Telescope and existing literature. Through this analysis, they were able to directly compare the H I gas fraction and SFE between spirals in S+S and S+E pairs. They found that, although the H I gas fraction is nearly identical in both pair types, the SFE of S+S pairs is ∼4.6 times higher. He et al. (2022) further investigated this finding by classifying SFGs according to their bulge-to-total ratio (B/T), utilizing the same data set of 88 close pairs as Cao et al. (2016). They ascertained that only SFGs in S+S pairs with less massive bulges (B/T < 0.3) exhibit significant sSFR enhancement, suggesting SFE enhancement is confined to disky SFGs within S+S pairs. Most recently, Pérez-Millán et al. (2023) discovered that the proximity and morphology of the nearest neighboring galaxies play an important role in shaping the current SF activity, even within cluster environments. Specifically, they found that late-type neighbors are more likely to enhance SF. Overall, these empirical observations provide strong support to the theoretical expectation that hydrodynamical effects exercise a critical influence on galaxy interactions, particularly in terms of how hot halo gas (e.g., Hwang et al. 2011; Moster et al. 2011, 2012; Wetzel et al. 2013; Hwang & Park 2015; Correa et al. 2018b; Rafieferantsoa et al. 2019) and cold disk gas (e.g., Renaud et al. 2015; Cao et al. 2016; Pan et al. 2018; Zuo et al. 2018; Moreno et al. 2019) engage with and contribute to these interactions.

Our first paper (Moon et al. 2019, hereafter Paper I), of the series named "Living with Neighbors" (An et al. 2019, 2021; Moon et al. 2021), undertook a comprehensive study of the hydrodynamical impacts of neighbor galaxies using an extensive sample of ∼7000 galaxy pairs, from SDSS Data Release 7 (DR7). Through a meticulous process of constructing isolated control samples, we established that neighbors with higher levels of SF and proximity ($\Delta R < 50\ h^{-1}$ kpc) can effectively increase sSFR. We interpreted this effect as a consequence of heightened collisions within the interstellar medium, in relation to the presence of star-forming neighbors. Additionally, our investigation revealed a contrasting phenomenon in galaxy pairs that have quiescent neighbors. We speculated that this behavior can be ascribed to the depletion of gas reservoirs through the process of ram pressure stripping and the cessation of gas accretion precipitated by the hot-gas halos of nearby quiescent neighbors. Overall, this result enabled us to present statistically compelling evidence in support of the influence of hydrodynamical mechanisms, along with tidal effects, on SF enhancement during galaxy–galaxy pair interactions.

However, a certain degree of unexplained variability persists. The relative orientation of two interacting galaxies can influence SF enhancement during their mutual interaction. But this effect remains not fully explored. In this study, we endeavor to observationally ascertain the dependency of interaction-induced SF on the spin–spin alignment (SSA) of galaxy pairs. Our exploration is anchored on two foundational pillars: (a) an innovative approach to measuring the SSA of galaxy pairs through the analysis of observed images, and (b) a meticulous examination of known hydrodynamic effects. In Section 2, drawing from Paper I, we identify S+S pairs from SDSS DR7 and assemble a corresponding control sample of isolated galaxies. In Section 3, we reevaluate the hydrodynamical influence on interaction-induced SF in the pairs, comparing the sSFRs of pairs with neighbors of relatively higher and lower star-forming activity. We then quantify the SSA for each pair and elucidate how interaction-induced SF is correlated with the respective spatial orientation. In Sections 4 and 5, we discuss and summarize our results. We adopt the standard cold dark matter cosmology with $\Omega_m = 0.3$, $\Omega_\Lambda = 0.7$, $H_0 = 100\,h$ km s$^{-1}$ Mpc$^{-1}$, and $h = 0.7$ throughout this paper.

## 2. Data and Methodology

### 2.1. Observational Data

Generally, our sample selection adheres to the methodology described in Paper I. The galaxies employed in this study are extracted from the SDSS DR7 database (Abazajian et al. 2009), utilizing SDSS CasJobs. For our interacting galaxy sample, we prioritize galaxies with a high-confidence spectroscopic redshift (SpecObjAll.zConf > 0.7) to mitigate the inclusion of spurious pairs. It is imperative to note that the galaxies chosen are mandatorily classified as "galaxy" both photometrically (PhotoObjAll.type=3) and spectroscopically (SpecObjAll.specClass=2). Please note that these selection criteria are in alignment with previous observational studies (e.g., Ellison et al. 2011, 2013; Scudder et al. 2012; Patton et al. 2013) that similarly explored the orbital extent of SF enhancement using a vast galaxy pair sample. We restrict our data set to galaxies within the redshift range of $0.02 < z < 0.06$. This specific range ensures accurate local density measurements, consistent with the comprehensive completeness criteria highlighted by Baldry et al. (2006). The intrinsic properties of the galaxies, including their stellar mass and disk inclination, are sourced from the catalogs described by Simard et al. (2011) and Mendel et al. (2014). These catalogs provide comprehensive photometric data for ~660,000 galaxies within the SDSS framework based on the broadband spectral energy distribution fitting with dusty models by Simard et al. (2011). It is important to note that our sample contains highly inclined (nearly edge-on) galaxies. For these galaxies, as Mendel et al. (2014) cautioned, discrepancies can arise between the total stellar mass and the sum of the decomposed bulge and disk stellar masses. To maintain accuracy in our analysis, we confine our selection to galaxies that match within a range of five standard deviations from the correlation between the total stellar mass and the aggregate of the bulge and disk masses.

The emission line fluxes and stellar kinematics are extracted from Sarzi et al. (2006) and Thomas et al. (2013). These measurements were obtained from SDSS spectra using the Gas and Absorption Line Fitting (GANDALF) code alongside the Penalized Pixel Fitting (pPXF) code, as detailed by Cappellari & Emsellem (2004). The catalog enumerates the emission line fluxes, specifically H$\alpha$, H$\beta$, [O III], and [N II]. To explore interaction-induced SF without contamination by emission lines of active galactic nuclei (AGNs), we segregate our sample into non-AGN and AGN host galaxies based on their distribution on the BPT diagram (Baldwin et al. 1981; Kauffmann et al. 2003). Our AGN sample is further refined to include only galaxies with H$\alpha$, H$\beta$, [O III], and [N II] lines exhibiting a signal-to-noise ratio >3. Galaxies surpassing the theoretical lower threshold for AGN host galaxies, as defined by Kauffmann et al. (2003), are categorized as AGN hosts, whereas galaxies either falling beneath this demarcation or absent from the BPT diagram due to feeble emission lines are designated as non-AGN. For the purposes of this study, only the latter category is utilized. The influence of galaxy–galaxy interactions on SF activity is conventionally anticipated to happen within the central region, specifically within a span of 1–3 kpc (e.g., Patton et al. 2011; Barrera-Ballesteros et al. 2015; Moreno et al. 2015; Garay-Solis et al. 2023). To scrutinize this central SF enhancement, we use SFR$_{\rm Fiber}$ and sSFR$_{\rm Fiber}$ from the MPA/JHU catalog (Brinchmann et al. 2004), which are determined through the reddening-corrected H$\alpha$ luminosity. Within the redshift range under consideration in this study ($0.02 < z < 0.06$), the 3″ fiber diameter captures a physical extent of 1–3 kpc.

### 2.2. Identification of Spiral+Spiral Pairs

We adopt the definition of galaxy pairs delineated by Paper I. Initially, for each galaxy in our sample, we identify the proximate candidates within a radial velocity interval of $|\Delta v| < 1000$ km s$^{-1}$. We consider only those neighboring galaxies whose stellar mass ratio relative to the target is higher than 1/10. Should a galaxy lack neighboring candidates within a projected separation $\Delta R < 200\ h^{-1}$ kpc, we categorize it as a "purely isolated galaxy" and reserve it for subsequent use as part of the control sample.

We construct our paired galaxy sample based on the following criteria:

1. *Projected physical separation*. The nearest neighbor must be located within a distance of $200\ h^{-1}$ kpc from the target galaxy. Additionally, we require a separation greater than the minimum value of 3″ on the sky to circumvent blending effects of nearby galactic lights.
2. *Radial velocity difference*. The nearest neighbor should exhibit a radial velocity that differs from the target's by no more than 300 km s$^{-1}$.
3. *Mass ratio*. In our quest to focus on galaxy pairs where the member galaxies possess roughly comparable masses, we restrict the nearest neighbor's stellar mass between 0.1 and 10 times that of the target galaxy. This criterion serves to minimize the influence of external perturbers.
4. *Survey boundary constraint*. Our analysis is confined to galaxies that maintain a separation from the survey boundary exceeding $4\ h^{-1}$ Mpc.

Our purpose is to estimate the misalignment angle between the spin vectors of the disks of the paired galaxies; this measurement is most reliable for S+S pairs. To distill our sample to only S+S pairs, we cross-reference it with the morphological classification data from Galaxy Zoo 2 (Willett et al. 2013). Based on its visual classification scheme, we select galaxies that are flagged as spirals requiring at least 80% of the volunteer votes to fall within the spiral category—"Clockwise" (Z-wise spiral galaxy), "Anticlockwise" (S-wise spiral galaxy), or "Spiral galaxy other" (e.g., edge-on galaxy)—after the debiasing procedure (see also Table 1 of Lintott et al. 2008). As detailed in Section 2.4, we compute the misalignment angle between spin vectors by considering both the position angle and the inclination of each disk. Consequently, our sample spans the range from highly inclined, nearly edge-on disks to face-on disks without any restricted range of apparent axis ratio. Upon cross-matching with our SDSS DR7 galaxy pair sample, we are left with 2898 galaxies in S+S pairs, wherein both interacting galaxies are classified as spiral galaxies with varying inclinations.

### 2.3. Relative SF Activity of the Nearest Neighbors

Adopting these methodologies, we divide our paired galaxy samples based on the sSFR of their neighboring galaxies, sSFR$_{\rm Nei}$. The left panel of Figure 1 illustrates the distribution of both sSFR$_{\rm Nei}$ and stellar mass for these neighbors. Traditionally, studies often bifurcate their samples based solely on sSFR$_{\rm Nei}$ (e.g., Cao et al. 2016; Paper I; He et al. 2022; Brown et al. 2023); however, there is a subtle inverse correlation: as the stellar mass of neighbors increases, sSFR$_{\rm Nei}$ decreases. To address this intrinsic bias, we segregate our sample according to the linear median distribution of sSFR$_{\rm Nei}$, represented by a dashed line in Figure 1. It is crucial to note that our study focuses exclusively on S+S pairs, meaning that the neighboring galaxies in our sample are spiral. Consequently, the sSFR$_{\rm Nei}$ distribution leans toward the star-forming sequence, especially with values of sSFR$_{\rm Nei} > 10^{-11}$ yr$^{-1}$. In light of this, neighbors with sSFR$_{\rm Nei}$ values above this median line are termed as "relatively higher SF" neighbors, while those with values beneath are termed "relatively lower SF" neighbors. For clarity, it is essential to understand that our term "relatively lower SF" deviates from the traditional "quiescent" descriptor frequently employed in previous studies.

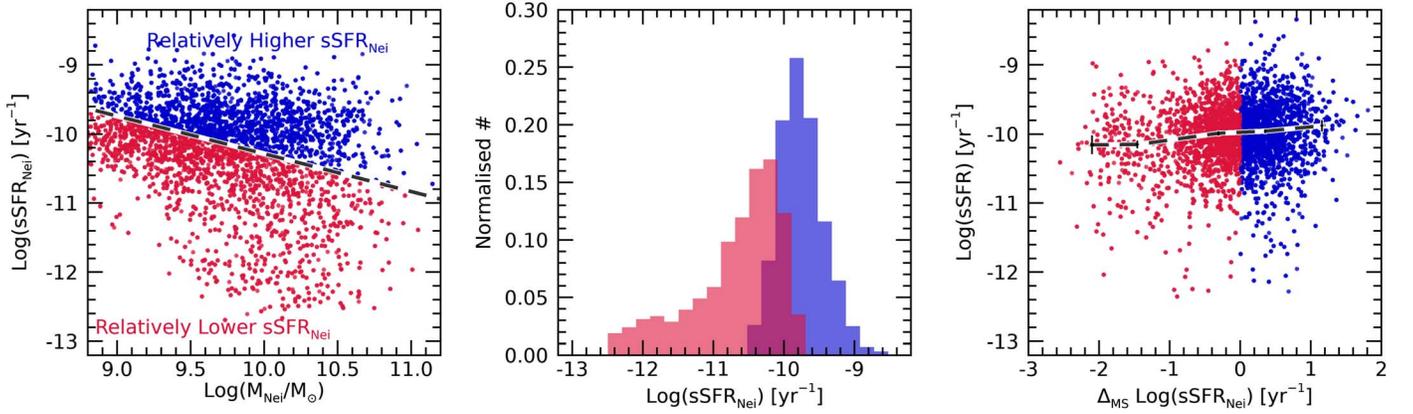

**Figure 1.** Left: The distribution of sSFR of the nearest neighboring galaxies, sSFR$_{\rm Nei}$, as a function of their stellar mass, $M_{\rm Nei}/M_\odot$. The black dashed line represents the division line, defined as the main sequence through linear median fitting of the distribution of all neighboring galaxies. In all panels, blue and red colors represent cases of relatively higher and lower SF in the neighbors, respectively. Middle: a normalized histogram for sSFR$_{\rm Nei}$. Right: The distribution of sSFR for target galaxies as a function of the deviation of sSFR$_{\rm Nei}$ from the main sequence, $\Delta_{\rm MS}{\rm Log(sSFR_{Nei})}$. The black dashed line represents the median distribution, with statistical errors via a jackknife resampling.

Figure 1 displays the distributions of the sSFR for both neighboring and target galaxies. We calculate the deviation of the sSFR of the neighbors from the linear median distribution of sSFR$_{\rm Nei}$, denoting it as $\Delta_{\rm MS}{\rm Log(sSFR_{Nei})}$. Notably, as illustrated in the right panel, the distribution of the sSFR of target galaxies as a function of $\Delta_{\rm MS}{\rm Log(sSFR_{Nei})}$ exhibits a subtle upward trend. This suggests that galaxies with neighbors of relatively higher SF activity tend to have higher SF activity themselves. This observation agrees with previous findings, including those presented in Paper I, which underscored the hydrodynamic effects exerted by neighboring galaxies. Note that our results remain consistent even when we adopt the conventional method of defining higher- and lower-SF neighbors using a simple horizontal dividing line at, e.g., Log(sSFR$_{\rm Nei}$) = −10.

### 2.4. Measurement of Spin–Spin Alignment

Adopting the pioneering technique developed by Lee & Pen (2002) and Lee (2011), we make an approximate estimation of the directions of the spin angular momentum vectors for the paired galaxies. This estimation is based on observed imagery and presumes a circular thin-disk approximation. To pinpoint the unit spin vector, $\hat{J}$, for each chosen disk galaxy, one requires the equatorial position ($\alpha$, $\delta$), position angle ($\psi_p$), and inclination ($i$). The spin vectors, when expressed in spherical coordinates as delineated by Lee & Erdogdu (2007), are represented by

$$\hat{J}_r = \cos i, \quad (1)$$
$$\hat{J}_\theta = \sqrt{1 - \cos^2 i} \, \sin \psi_p, \quad (2)$$
$$\hat{J}_\phi = \sqrt{1 - \cos^2 i} \, \cos \psi_p, \quad (3)$$

where $\hat{J}_r$, $\hat{J}_\theta$, and $\hat{J}_\phi$ represent the radial, polar, and azimuthal components of $\hat{J}$, respectively. It is important to note the degree of ambiguity with the direction of $\hat{J}$. Based solely on the observed image, it is challenging to discern if the target disk rotates clockwise or counterclockwise (Pen et al. 2000; Lee 2011; Koo & Lee 2018). Consequently, when translating $\hat{J}$ into Cartesian coordinates, $\hat{J} = (\hat{J}_x, \hat{J}_y, \hat{J}_z)$, the unit spin vector for each primary and secondary galaxy introduces a twofold degree of freedom as follows:

$$\hat{J}_{x\pm} = \pm \hat{J}_r \sin\theta \cos\phi + \hat{J}_\theta \cos\theta \cos\phi - \hat{J}_\phi \sin\phi, \quad (4)$$
$$\hat{J}_{y\pm} = \pm \hat{J}_r \sin\theta \sin\phi + \hat{J}_\theta \cos\theta \sin\phi + \hat{J}_\phi \cos\phi, \quad (5)$$
$$\hat{J}_{z\pm} = \pm \hat{J}_r \cos\theta - \hat{J}_\theta \sin\theta, \quad (6)$$

where $\theta = \pi/2 - \delta$ and $\phi = \alpha$ (see also Pen et al. 2000; Lee & Pen 2002; Lee & Erdogdu 2007; Lee 2011; Koo & Lee 2018; Lee & Moon 2023). This ambiguity culminates in 4 distinct degrees of freedom (denoted as $\xi_1$, $\xi_2$, $\xi_3$, and $\xi_4$) as shown below:

$$\cos \xi_1 = |\hat{J}_{p+} \cdot \hat{J}_{s+}|, \quad (7)$$
$$\cos \xi_2 = |\hat{J}_{p+} \cdot \hat{J}_{s-}|, \quad (8)$$
$$\cos \xi_3 = |\hat{J}_{p-} \cdot \hat{J}_{s+}|, \quad (9)$$
$$\cos \xi_4 = |\hat{J}_{p-} \cdot \hat{J}_{s-}|, \quad (10)$$

where $\hat{J}_{p\pm}$ and $\hat{J}_{s\pm}$ are the unit spin vector of the primary and secondary galaxy in each twofold degree of freedom, respectively, as graphically represented in Figure 2. We seek to assess the interaction-induced SF enhancement as a continuous function of the misalignment angle between the spin vectors of the paired galaxies. In line with the methodology proposed by Koo & Lee (2018), we simultaneously incorporate values from all four degrees of degeneracy. Given this definition, we obtain $4N$ realizations of $\cos \xi$ from a total of $N$ S+S pairs. Figure 3 shows post-stamp examples of multiband SDSS images of S+S pairs, ranging from perpendicular configurations ($\cos \xi \sim 0$) to well-aligned systems ($\cos \xi > 0.7$) with varying projected separations. Figure 4 presents the distribution of misalignment angles between the two spin vectors, represented as $\cos \xi$ from our comprehensive S+S pair samples. There is a pronounced trend toward increasing alignment as pair systems become more coherently oriented, irrespective of the sSFR of the adjacent galaxies. This suggests that the orientations of the spin axes for the primary and secondary members within S+S pairs are not arbitrary, but rather exhibit a tendency toward mutual alignment. This observation aligns well with previous observational statistics and theoretical predictions (Pestaña & Cabrera 2004; Slosar et al. 2009; Koo & Lee 2018).

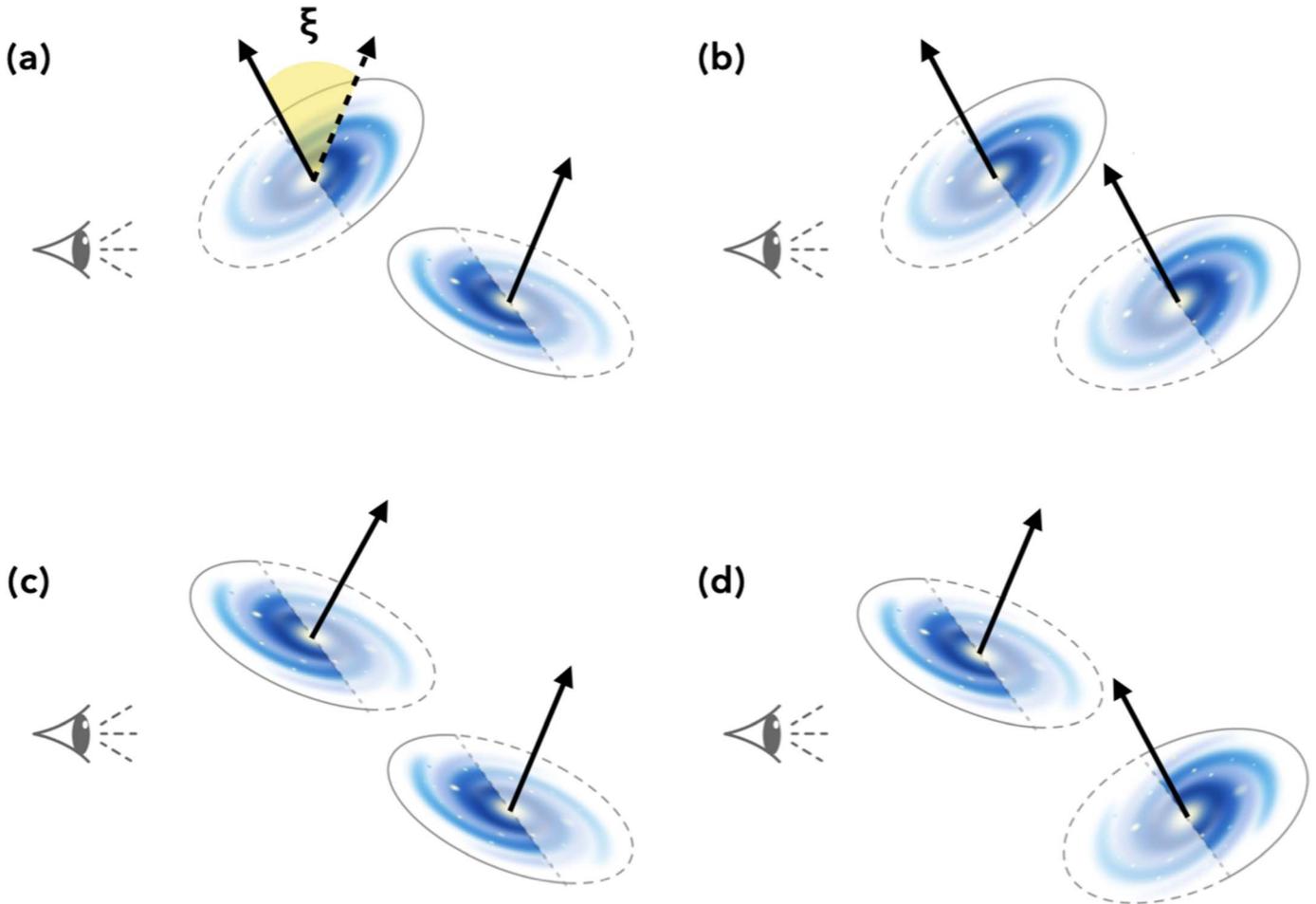

**Figure 2.** Graphical depiction of the fourfold degree of freedom in relative orientation between the spin axes of member galaxies in S+S pairs. Each degeneracy arises from the combined freedom in inclination and the positive/negative spin orientation of each disk. In all panels, black solid arrows indicate the direction of the spin axis of each disk. To visually depict inclined disks, the areas above and below the plane of sight are distinguished by darker and paler shades, respectively. In panel (a), the misalignment angle, $\xi$, between the two spin axes illustrates the degree of SSA.

### 2.5. Control Sample

To investigate the influence of galaxy–galaxy interactions on SF activity, the meticulous construction of an isolated control sample with precision is imperative. As in Paper I, we demonstrate a comprehensive methodology for curating this control sample, ensuring that it is free from intrinsic biases related to redshift, stellar mass, or local density. Moreover, to mitigate the influence of disk inclination on sSFR measurements, our control sample accounts for the observed distribution of disk inclinations of the target galaxies. Building on this framework, the control sample is derived from the isolated galaxy sample delineated in Section 2.2. By definition, these isolated galaxies are devoid of any proximate neighbors within a projected span of $\Delta R < 200 \, h^{-1}$ kpc and a radial velocity difference range of $|\Delta v| < 1000$ km s$^{-1}$. For each paired galaxy, we iteratively and randomly select an isolated galaxy 100 times that aligns within a deviation of $\pm 0.005$ dex in redshift, $\pm 0.1$ dex in stellar mass, $\pm 0.1$ dex in local density, and $\pm 10°$ in disk inclination. As elaborated in Paper I, we ascertain the local density for this control sample through the adaptive kernel technique, as expounded by Silverman (1986; see also Ferdosi et al. 2011; Darvish et al. 2015 and references therein).

The top panels of Figure 5 depict the distribution of redshift, stellar mass, local density, and disk inclination of the uncontrolled isolated and pair samples. The distribution of redshift reveals that paired galaxies are more prevalent at lower redshift. Further, analysis of other properties shows that paired galaxies generally have a lower stellar mass and reside in denser environments. These differences in distributions largely stem from selection biases. The SDSS fiber collision and blending effect makes the detection of close pair systems at high redshifts challenging. Given the predominant population of low-mass galaxies in the Universe, it becomes more likely for lower-mass galaxies to have neighbors of comparable mass. Moreover, it is intuitively expected that paired galaxies would predominantly be found in denser environments. Although the distribution of disk inclinations in uncontrolled isolated galaxies does not significantly differ from that of paired galaxies, we modify our control sample matching it with the disk inclination distribution of the paired sample. This approach is adopted to minimize the impact of intrinsic dust attenuation on the measurement of sSFR. Once these intrinsic biases are accounted for in the construction of the control sample, the distribution becomes nearly identical to that of the paired galaxy sample, as illustrated in the bottom panels of Figure 5. Following this establishment of a control sample, we identify 2783 S+S pair galaxies that could be matched with isolated counterparts, whereas some pairs could not be appropriately matched within the range of their intrinsic properties. Consequently, the total number of the control sample amounts to 278,300 through 100 iterations of matching.

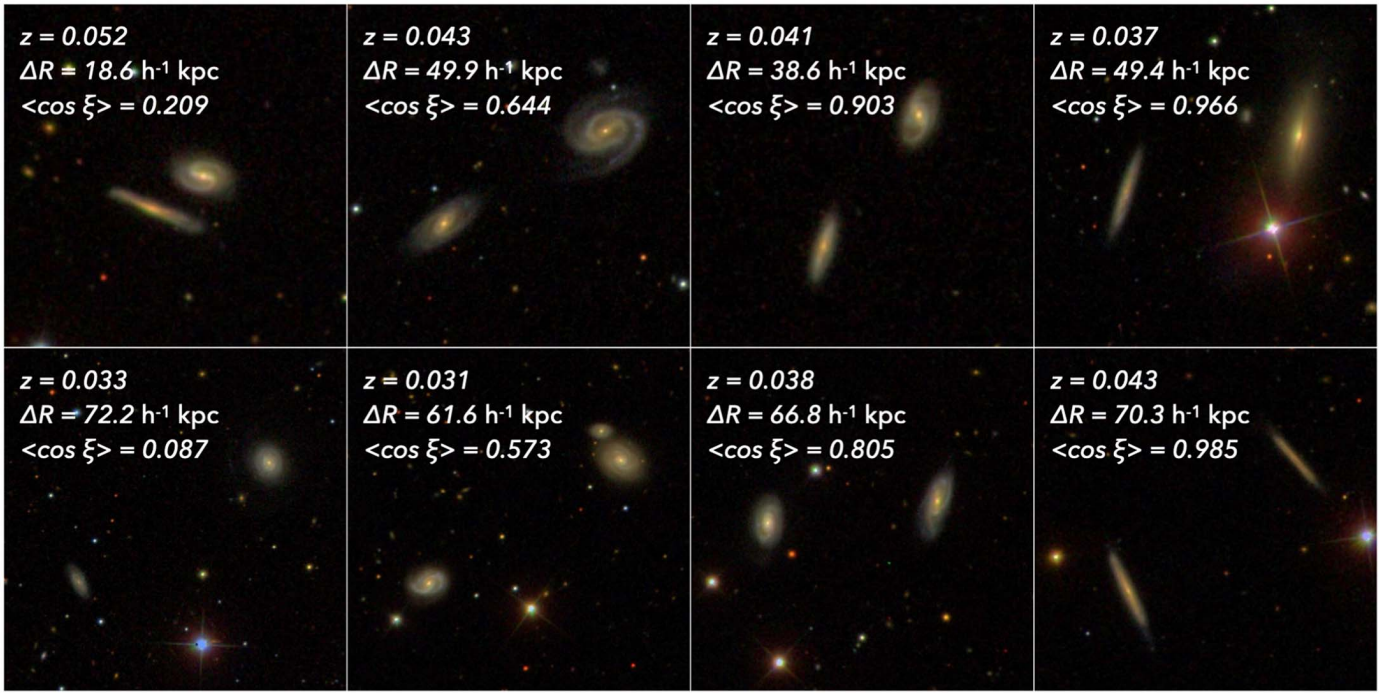

**Figure 3.** An example of multiband SDSS images featuring S+S pair galaxies identified through our automatic classification. For comparison purposes, we provide information on the redshift, the projected separation between the two members, and the misalignment angle between their respective spin axes. To simplify matters, we present the median value of the misalignment angle, which has four possible configurations for each pair. The projected separation increases from top to bottom, and the configuration becomes more aligned from left to right.

## 3. Results

### 3.1. Revisiting Tidal and Hydrodynamical Effects: Closer and Bluer Neighbors Enhance SF

We revisit the main conclusion of Paper I: paired galaxies exhibit augmented sSFRs as compared to an isolated control sample, particularly at closer projected separations ($\Delta R < 80\,h^{-1}$ kpc), extending to farther distances ($\Delta R \sim 200\,h^{-1}$ kpc) when star-forming neighboring galaxies are present. The left panels of Figure 6 show the sSFR of target galaxies as a function of projected separation for target galaxies with relatively higher SF neighbors (top panel), and for those with relatively lower SF neighbors (bottom panel). Note that an explanation of the right panels of Figure 6 is provided in Section 3.2. We define interaction-induced SF as the ratio of the sSFR of target galaxies within S+S pairs to that of a corresponding isolated control sample, represented as $\Delta\mathrm{sSFR} \equiv \mathrm{sSFR}/\mathrm{sSFR}_{\mathrm{Control}}$, and illustrate $\Delta\mathrm{sSFR}$ as a function of the projected separation between the paired galaxies.

The influence of proximate neighbors on $\Delta\mathrm{sSFR}$ is vigorous when those neighbors demonstrate relatively higher SF activity, as depicted in the top middle panel of Figure 6. This finding aligns with previous research suggesting that more active, bluer, and gas-rich neighbors induce more SF through hydrodynamical processes, such as increased SFE and enhanced gas accretion during interactions. Moreover, it is compelling that our distinction between "relatively higher SF" and "relatively lower SF" neighbors, albeit less rigid than classifications in earlier studies, still presents clear differences. This suggests that hydrodynamical influences on interaction-induced SF operate as a continuous function of the neighbor's sSFR, evolving in a systematic trend.

### 3.2. Augmentation of Interaction-induced SF in Well-aligned Pair Systems

In the left panels of Figure 6, the color gradient represents $\cos \xi$, utilizing the well-established local regression approach from Cappellari et al. (2013). In the top panel, when target galaxies are paired with relatively higher SF neighbors, there is a discernible systematic increase in $\Delta\mathrm{sSFR}$ as $\cos \xi$ approaches 1, particularly at $\Delta R < 80\,h^{-1}$ kpc. In contrast, the bottom panel does not exhibit any systematic trend. Upon dividing our sample into less aligned ($\cos \xi \leqslant 0.7$) and well-aligned systems ($\cos \xi > 0.7$), as depicted in the right panels of Figure 6, the dependency on SSA becomes more apparent. Intriguingly, we observe a consistent twofold increase in $\Delta\mathrm{sSFR}$, ranging from less aligned configurations ($\cos \xi \leqslant 0.7$) to well-aligned systems ($\cos \xi > 0.7$), specifically at closer projected separation ($\Delta R < 80\,h^{-1}$ kpc). In both cases of "relatively higher" and "relatively lower" SF neighbors, a consistent pattern of SF enhancement is observed in well-aligned systems. Notably, the distinction between well-aligned

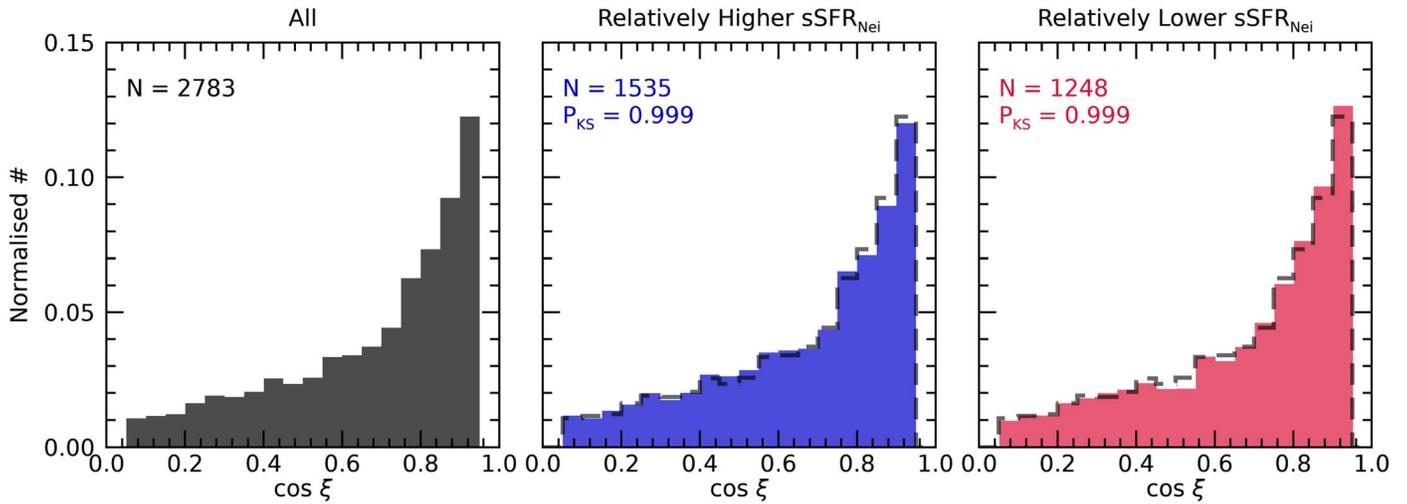

**Figure 4.** Normalized distribution of cos ξ for all target galaxies (left, in black), galaxies with relatively higher SF neighbors (middle, in blue), and those with relatively lower SF neighbors (right, in red). For comparison, the distribution of all target galaxies is depicted as a black dashed line in both the middle and right panels. Each p-value by the Kolmogorov–Smirnov test is given on each panel.

and less aligned systems is more pronounced in galaxies with "relatively higher SF" neighbors.

Figure 7 allows us to examine the SF enhancement as a function of the SSA more directly. In the left panels, the color gradient represents cos ξ. A distinct trend is evident: the sSFR of target galaxies increases as the projected separation between the paired members decreases. This is illustrated by the color gradient transitioning from red to blue. Moreover, as illustrated in the middle and right panels of Figure 7, upon contrasting the distribution of ΔsSFR against cos ξ, one finds that there is a discernible trend: in S+S pairs, the sSFR of target galaxies tends to rise in tandem with growing cos ξ, particularly at closer projected separations ($\Delta R \leqslant 80\ h^{-1}$ kpc). Interestingly, when target galaxies are paired with neighbors characterized by "relatively higher SF," the SSA-driven interaction-induced SF is more pronounced. However, for those target galaxies paired with "relatively lower SF" neighbors, the variation in ΔsSFR remains largely consistent across the range of SSA signals. This result aligns with previous studies that suggested the pivotal role of hydrodynamic effects, as evidenced by enhanced SF in the presence of star-forming, blue, and gas-rich neighbors. Consequently, the augmented ΔsSFR observed in well-aligned systems in this study plausibly arises from these hydrodynamic effects.

## 4. Discussion

Our previous work (Paper I) showed that paired galaxies experience enhanced interaction-induced SF in the presence of star-forming neighbors. Here, we delve deeper, suggesting that SSA, a factor frequently bypassed in earlier research, may play an important role in modulating interaction-induced SF. Notably, our results indicate that interaction-induced SF increases as alignment changes from perpendicular to parallel, especially when neighboring galaxies exhibit relatively higher sSFR.

Before investigating the exact mechanisms underlying the correlation between SSA and SF enhancement, it is essential to ascertain whether interaction-induced SF truly hinges on SSA. A challenge arises from the possibility that the proximity between paired galaxies not only enhances SF but also causes their alignment.

### 4.1. Does a Well-aligned Configuration Result from Interactions That Enhance SF?

The relative orientation of pair galaxies is primarily determined by the acquisition of angular momentum in the protogalaxies, which are affected by adjacent tidal fields. Traditionally, both observational (e.g., Lee & Pen 2001, 2002, 2007, 2008; Brown et al. 2002; Faltenbacher et al. 2007; Hirata et al. 2007; Lee & Erdogdu 2007; Okumura et al. 2009; Slosar et al. 2009; Lee 2011; Hung & Ebeling 2012; Li et al. 2013; Zhang et al. 2013; Krolewski et al. 2019; Kraljic et al. 2021; Shamir 2022) and computational studies (e.g., Porciani et al. 2002; Bailin & Steinmetz 2005; Bailin et al. 2005; Faltenbacher et al. 2008; Okumura & Jing 2009; Trowland et al. 2013; Prieto et al. 2015; González et al. 2017; Lee 2019; Shi et al. 2021) have focused on understanding the alignment between the spin axes of galaxies and the large-scale filaments nearby. Spin vectors of spiral galaxies are commonly found to be parallel to these large-scale filaments. This observed large-scale alignment is believed to be driven by the combined influence of surrounding tidal torque fields (Lee & Pen 2002, 2008; Porciani et al. 2002; Hirata et al. 2007; Lee & Erdogdu 2007; Faltenbacher et al. 2008; Lee 2011, 2019; Zhang et al. 2013; Kraljic et al. 2021) and/or accretion along filaments (Bailin & Steinmetz 2005; Trowland et al. 2013; Prieto et al. 2015; Krolewski et al. 2019). Given these findings, it is also reasonable to infer that neighboring galaxies, which are subject to the same large-scale structure, might exhibit spin orientations that are similarly aligned.

In recent decades, an emerging body of research has sought to more directly investigate SSA; however, the evidence for pronounced alignment in closely proximate galaxy pairs remains inconclusive (Joachimi & Schneider 2010; Andrae & Jahnke 2011; Lee 2011; Buxton & Ryden 2012; Chisari et al. 2015; Sifón et al. 2015; Koo & Lee 2018; Krolewski et al. 2019). Such alignment signals also seem to be subtly influenced by various intrinsic properties of the galaxies, including the halo mass (Li et al. 2013; Trowland et al. 2013; Xia et al. 2017), bulge mass (Barsanti et al. 2022), SF activity (Jimenez et al. 2010; Shi et al. 2021), and spiral morphology (Hu et al. 1995; Lee 2011; Koo & Lee 2018). For example,

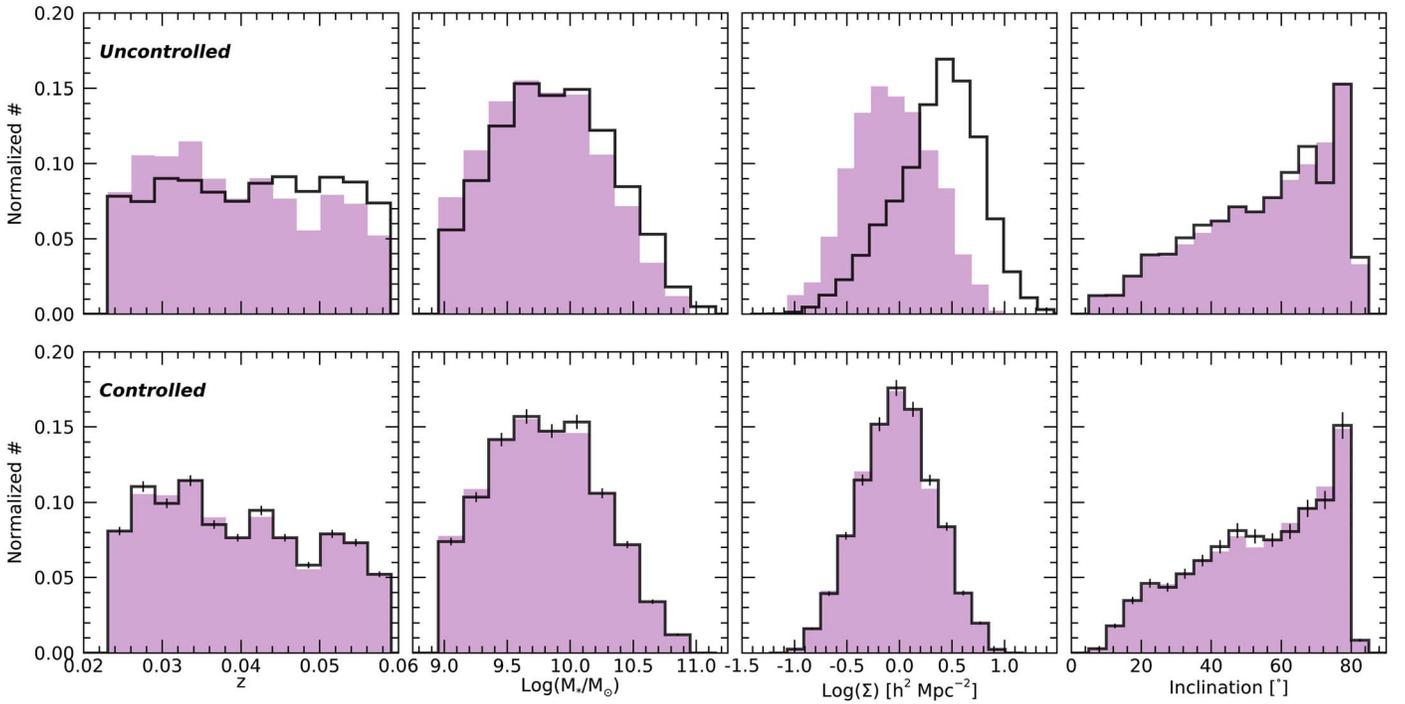

**Figure 5.** Top: the normalized distribution of redshift (first column), stellar mass (second), local density (third), and inclination (fourth) for uncontrolled isolated galaxies (depicted by thick empty histograms) and S+S pair galaxies (represented by filled purple histograms). Bottom: same as the top row, but for the controlled isolated galaxies. Error bars indicate the standard deviation obtained from 100 random selections within the control sample.

Koo & Lee (2018) showed that intrinsic spin alignments in isolated spiral pairs are marginally more prominent for earlier-type spiral galaxies (Sa–Sb) than for later types (Sc–Sd). Conversely, the same analysis conducted using isolated dark matter subhalo pairs from the EAGLE hydrodynamic simulations was unable to reproduce the observed correlation of strengthened alignment in close pairs. They hypothesized that this tension between observation and simulation could primarily originate from the estimation methods of spin vectors in galaxies from the EAGLE simulation, which were determined by the dark matter halo, whereas stellar disks are utilized in observations. Moreover, using the Horizon hydrodynamical simulation, Soussana et al. (2020) cautioned that AGN feedback could modify both the selection of galaxy pairs and their observed alignments.

Interestingly, a pioneering study by Jimenez et al. (2010) proposed that the spin alignment signal of neighboring spiral galaxies is correlated with their SF history. Utilizing spin directions from SDSS galaxies provided by the Galaxy Zoo project and adopting an automated algorithm, they demonstrated that galaxies that primarily formed their stars early ($z > 2$) exhibit coherent spin directions across scales of $\sim 10 \, h^{-1}$ Mpc. In contrast, galaxies with significant recent SF lack such systematic large-scale spin alignments. The authors postulated that galaxies with older stellar populations predominantly formed within filaments in the early Universe. Meanwhile, galaxies with higher recent SF might have experienced mergers that abruptly altered their spin directions. These findings seem to contrast with our observations, which indicate enhanced interaction-induced SF in well-aligned galaxy pairs. Nevertheless, a significant distinction exists in the methodologies. Jimenez et al. (2010) estimated the average spin chirality across numerous neighboring galaxies over extensive regions (2–50 $h^{-1}$ Mpc) and exclusively used nearly face-on spiral galaxies. In contrast, our approach uses a direct measurement of SSA for each galaxy pair without accounting for their inclination.

Whether SSA depends on the proximity of the paired galaxies is still under debate. A meticulous examination is required to confirm whether the more significant interaction-induced SF for well-aligned systems is indeed a result of the alignment itself. Pursuant to these considerations, we directly examine the correlation between $\cos \xi$ and the projected separation between pair members while keeping other parameters constant. Figure 8 demonstrates the absence of a systematic trend in $\cos \xi$ with respect to the projected separation in our sample. As in Figures 6 and 7, the color gradient in the left panels of Figure 8 depicts the $\Delta$sSFR of the target galaxies. The strength of the SSA signal appears constant regardless of the proximity between the paired galaxies. The only discernible trend is the augmentation of interaction-induced SF in more aligned pair systems, specifically those in close proximity ($\Delta R < 80 \, h^{-1}$ kpc) to neighbors with relatively higher SF. Additionally, there is a subtle upward trend in $\Delta$sSFR toward more aligned configurations ($\cos \xi = 1$) exclusive to galaxies paired with higher-SF neighbors.

Drawing from previous studies that suggested the orientation of spin vectors of galaxies is influenced by tidal torque fields from nearby large-scale filaments, we examine SSA variations in relation to filaments. We match our data set with a large-scale structure catalog from Tempel et al. (2014), which captures filamentary patterns covering $\sim 40\%$ of the SDSS galaxies. In Figure 9, no significant systematic trend is discerned between $\cos \xi$ and the distance from the nearest filament structures. This suggests that the traditionally observed coherent intrinsic spin alignments of spiral galaxies, attributed to tidal torque fields, might emerge primarily in large-scale statistics ($>1 \, h^{-1}$ Mpc). It is important to highlight that our focus is exclusively on the SSA between paired galaxies within smaller scales ($\Delta R < 200\, h^{-1}$ kpc).

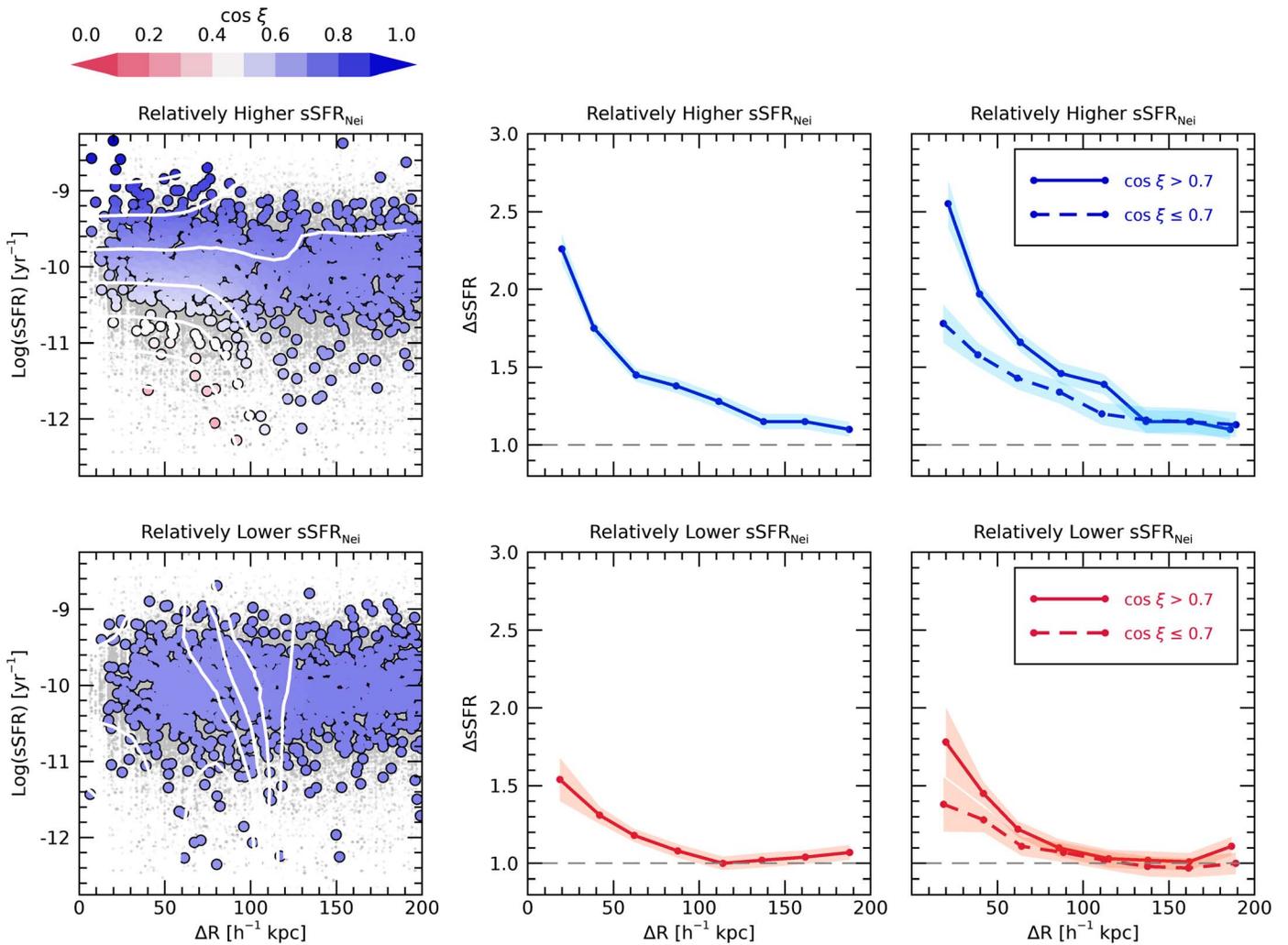

**Figure 6.** The distribution of the projected separation vs. the SF activity of paired galaxies. The upper and lower panels display results for target galaxies with relatively higher and relatively lower SF neighboring galaxies, respectively. Left: The distribution of sSFR for paired target galaxies (colored circles) and an isolated control sample (gray dots) as a function of the projected separation from the nearest neighbors. We apply the local regression scheme of Cappellari et al. (2013). The color gradient, shifting from redder to bluer, corresponds to increasingly aligned SSA configurations. Contour lines in white are presented at intervals of $0.5\sigma$ based on the distribution of $\cos \xi$. Middle: The interaction-induced SF enhancement, $\Delta$sSFR, defined as the ratio between the sSFR of the target galaxy and that of the matched control sample, plotted against the projected separation from the nearest neighbors. Colored bands indicate the range of statistical errors via a jackknife resampling. The dashed gray horizontal line at $\Delta$sSFR = 1 indicates where interaction-induced SF enhancement is absent. Right: Similar to the middle panel, but categorized based on the misalignment angle between the spin axes of the two pair members. Solid and dashed lines represent well-aligned ($\cos \xi > 0.7$) and less aligned ($\cos \xi \leqslant 0.7$) cases, respectively.

In essence, the SSA signal we derive does not seem to be heavily influenced by the proximity of the paired galaxies or the surrounding large-scale structures. This leads us to exclude the notion that a well-aligned configuration and SF are outcomes of the same interaction. In this regard, we further probe the mechanisms governing the relationship between SSA and interaction-induced SF in the next section.

### 4.2. How Does Alignment Enhance SF during Interaction?

In this section, we explore the potential mechanisms underlying the observed correlation between SSA and interaction-induced SF. Our findings reveal the following: (a) well-aligned pair systems display enhanced SF activity as compared to less aligned configurations, and (b) the SF activity in well-aligned systems increases in the presence of relatively higher star-forming activity neighbors. Drawing on the primary conclusions from Paper I, we suggest that (a) a more pronounced hydrodynamic friction via ram pressure is at work in better-aligned configurations (Section 4.2.1) and (b) a higher probability of prograde orbits in well-aligned neighboring galaxy pairs (Section 4.2.2) plausibly explains our new findings in this study.

#### 4.2.1. Hydrodynamic Friction by Ram Pressure

The augmented SF activity in well-aligned systems may be instigated by dynamical friction, which potentially promotes further gas inflow (e.g., Van Wassenhove et al. 2014; Capelo et al. 2015; Renaud et al. 2015; Capelo & Dotti 2017; Blumenthal & Barnes 2018). For example, Capelo & Dotti (2017) conducted simulations to examine the variation in ram pressure significance based on the relative orientation of two interacting galaxies. Their simulations indicate that during coplanar, prograde mergers, ram pressure shocks often induce abrupt changes in the angular momentum of gas, especially within the central 3 kpc, irrespective of other diverse impact parameters. This effect is usually observed when the galaxies almost entirely overlap, resulting in the gas disk undergoing intense ram pressure and becoming dissipative. The inflow of gas, driven by hydrodynamic processes, can adequately fuel a starburst event (McDermid et al. 2006; Blumenthal & Barnes 2018). On the other hand, mergers with greater inclination display minimal alterations in the spin vectors of gas components. In such configurations, only a minor fraction of the gas in each disk undergoes hydrodynamic effects, attenuating the influence of ram pressure shocks.

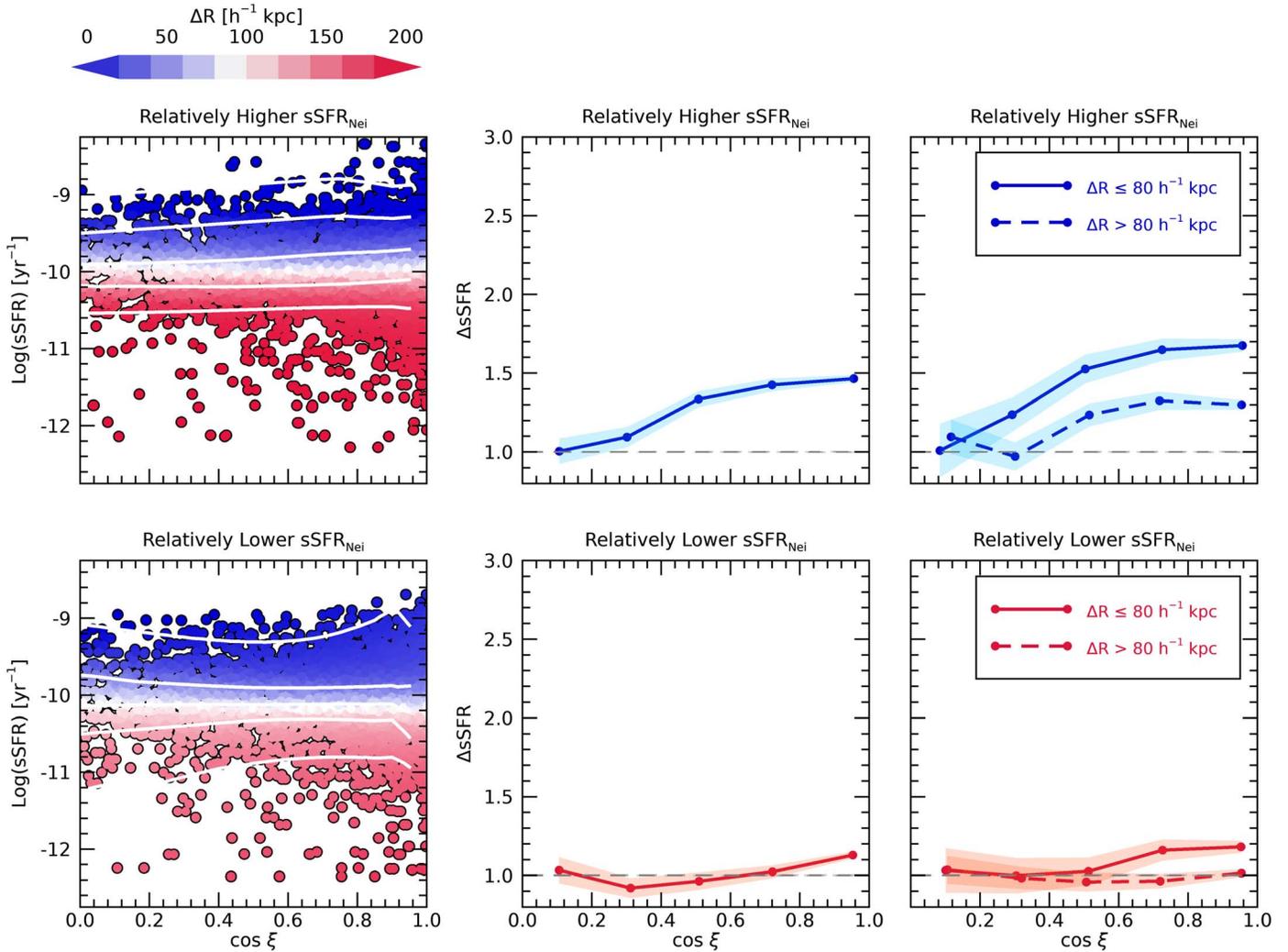

**Figure 7.** The distribution of spin alignments vs. SF activity of the paired galaxies. The upper and lower panels display results for target galaxies with relatively higher and relatively lower SF neighboring galaxies, respectively. Left: The distribution of sSFR for paired target galaxies as a function of the misalignment angle between the spin axes of the pair members, $\cos \xi$. We apply the local regression scheme of Cappellari et al. (2013). The color gradient, shifting from bluer to redder, corresponds to wider separation. Contour lines in white are presented at intervals of $0.5\sigma$ based on the distribution of $\Delta R$. Middle: The interaction-induced SF enhancement, $\Delta$sSFR, defined as the ratio between the sSFR of the target galaxy and that of the matched control sample, plotted against $\cos \xi$. Colored bands indicate the range of statistical errors via a jackknife resampling. The dashed gray horizontal line at $\Delta$sSFR = 1 indicates where interaction-induced SF enhancement is absent. Right: Similar to the middle panel, but categorized based on the projected separation from the nearest neighbors. Solid and dashed lines represent closer ($\Delta R \leqslant 80\ h^{-1}$ kpc) and extended ($\Delta R > 80\ h^{-1}$ kpc) cases, respectively.

Interestingly, hydrodynamic friction appears to favor the formation of well-aligned and circular orbits, especially in gas-rich interactions. When galaxies have initially anti-aligned angular momentum directions, they tend to align; conversely, if they begin in alignment, their orbits tend to circularize through hydrodynamic friction (Barnes 2002; Dotti et al. 2006; Mayer et al. 2007; Callegari et al. 2011; Capelo & Dotti 2017). Bonetti et al. (2020) suggested that this drift toward circular corotating orbits may influence the spin orientation distribution of disk galaxies, potentially resulting in a high frequency of corotating interactions. This notion seems to be in line with our observations, hinting at possible hydrodynamic effects in well-aligned galaxy pairs. It is important to mention, however, that these prior studies primarily focused on the change in angular momentum itself and did not delve deeply into the evolution of SF history.

There are relatively few theoretical studies that have delved into the influence of spin orientation on SF evolution during mergers, providing potential insights relevant to our findings. For instance, Rodríguez et al. (2022) demonstrated that the most significant gas removal events occur when the satellite galaxy's disk is perpendicular to its direction of motion (consistent with our definition of well-aligned disks for primary and satellite galaxies). Their simulations indicated that the ram pressure effect is more pronounced, leading to greater SF enhancement, in these specific configurations. They also posited that substantial alterations in disk orientation are infrequent during interactions. In their simulations, all physical processes, encompassing ram pressure, tidal torque, and SF activity, attain their peak values synchronously. This happens because the maximum ram pressure coincides with periods when the surrounding gas is densest and exhibits the highest relative velocity in relation to the central galaxy. However, these studies have analyzed results from models that alter not just the spin orientation but also other independent orbital parameters, such as the impact parameter and relative velocity. Consequently, it is challenging to isolate and understand the exclusive impact of relative spin orientation on SF history. As such, there is still a need for more comprehensive hydrodynamical simulations, taking into account both varying spin orientations and the amount of available gas.

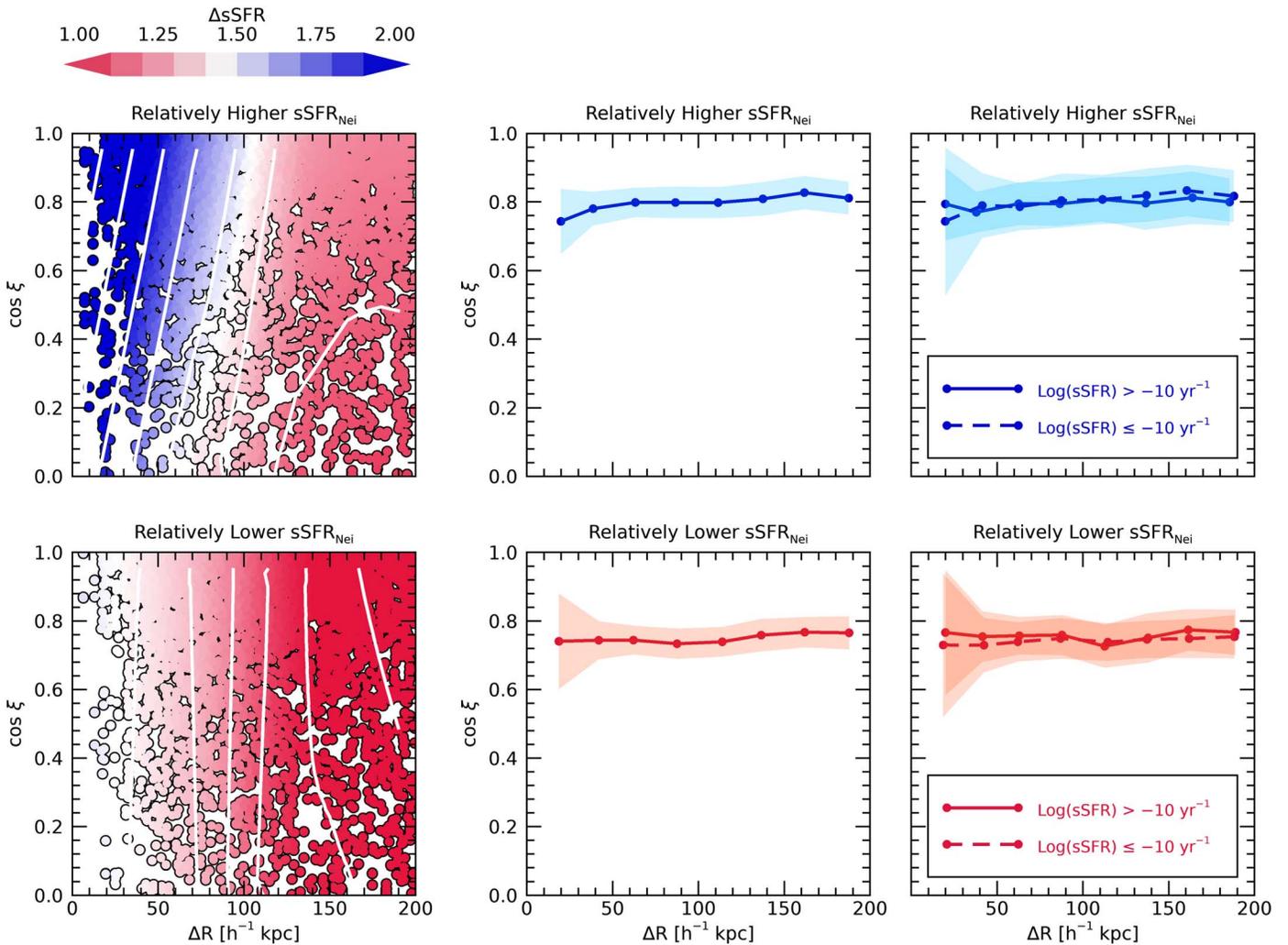

**Figure 8.** The distribution of projected separation vs. spin alignment of paired galaxies. The upper and lower panels display results for target galaxies with relatively higher and relatively lower SF neighboring galaxies, respectively. Left: The distribution of the misalignment angle between the spin axes of the pair members, $\cos \xi$, as a function of projected separation from the nearest neighbors. We apply the local regression scheme of Cappellari et al. (2013). The color gradient, shifting from redder to bluer, corresponds to higher interaction-induced SF enhancement. Contour lines in white are presented at intervals of $0.5\sigma$ based on the distribution of $\Delta$sSFR. Middle: The $\cos \xi$ value as a function of projected separation from the nearest neighbors. Colored bands indicate the range of statistical errors via a jackknife resampling. Right: Similar to the middle panel, but categorized based on the sSFR of the target galaxies. Solid and dashed lines represent higher-SF (sSFR > $10^{-10}$ yr$^{-1}$) and lower-SF (sSFR $\leqslant 10^{-10}$ yr$^{-1}$) cases, respectively.

#### 4.2.2. A Higher Probability of Prograde Orbits

SSA reflects the spin of each galaxy and may be modulated by the orientation between the spin axis and the orbital motions. Interactions with neighboring galaxies can concurrently shape the spin orientation and orbital trajectory of the galaxies (Capelo & Dotti 2017; L'Huillier et al. 2017; Choi et al. 2018; Lee et al. 2018b, 2019; An et al. 2021; Moon et al. 2021; Mai et al. 2022). Typically, the spins of two interacting galaxies in a pair exhibit a preferential alignment, often in the same direction (Pestaña & Cabrera 2004; Cervantes-Sodi et al. 2010; Mesa et al. 2014; Koo & Lee 2018; Lee et al. 2018a). This prevalent SSA in galaxy pairs is commonly attributed to the reciprocal interactions between two orbiting galaxies. Given these results, it is plausible to infer that the SSA of galaxy pairs reflects their relative orbital geometry—the spin–orbit alignment—which is well established in our series papers (An et al. 2019, 2021; Moon et al. 2021).

While the majority of prior studies do not heavily favor the direct influence of SSA on interaction-induced SF, recent advances in hydrodynamic simulations suggest that the effect of a well-aligned orbital geometry is as crucial to increasing SF activity as the available gas content in galaxies (e.g.,

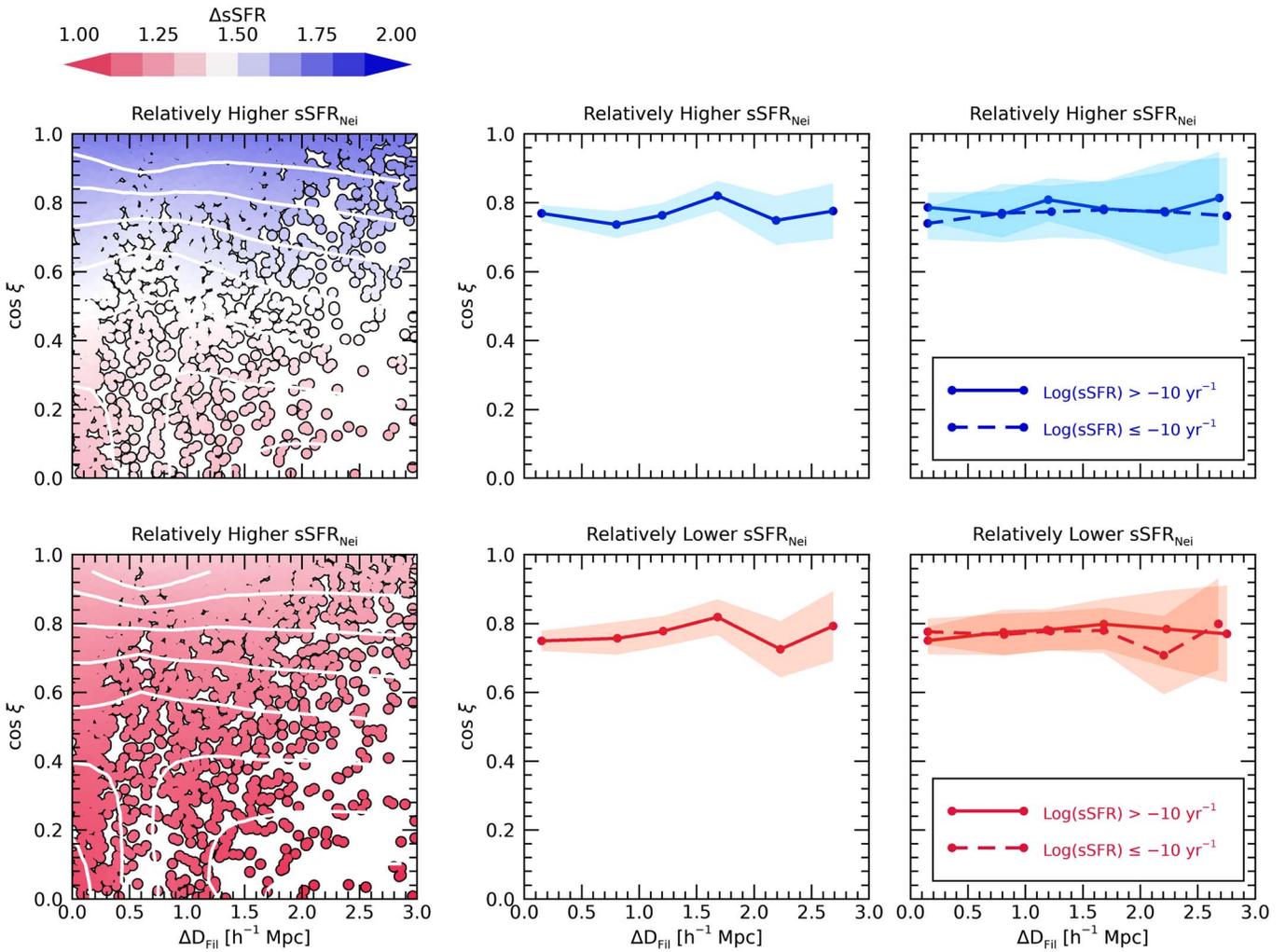

**Figure 9.** The same as Figure 8 but for the distribution of distance to the closest large-scale filaments, $\Delta D_{\mathrm{Fil}}$, vs. spin alignment of paired galaxies.

Mihos et al. 1992; Mihos & Hernquist 1996; Perez et al. 2005, 2006; Cox et al. 2008; Patton et al. 2013). For example, Cox et al. (2008) contended that direct (or coplanar) encounters induce the most intense starburst phase, especially when the primary disk lacks a bulge, leading to pronounced tidal disturbances. Simulations by Di Matteo et al. (2008) showed that perfect coplanar encounters bolster the SFR significantly and lead to rapid dissipation of gas angular momentum. These simulations revealed coplanar mergers to have an SFR peak nearly double that of polar encounters, with a marked decline in interaction-induced SFR as orbital inclination changes from 45° to 90°. In a similar vein, Park et al. (2017) used systematic simulations with varying orbital inclinations and observed that both prograde-orbit (0°) and retrograde-orbit (180°) mergers result in an enhanced SFR with respect to other inclinations. They also highlighted differences in peak SF over time between prograde and retrograde interactions: prograde encounters induce a delayed, sustained increase in SFR, whereas retrograde mergers trigger an immediate starburst at first passage during interaction.

According to these studies, flyby interactions, particularly those including prograde disks, induce a global starburst that is 2–3 times stronger compared to those induced by interactions with retrograde or highly inclined orientations (Mihos et al. 1992; Patton et al. 2013). The most potent gas inflow, leading to starburst, is observed in coplanar encounters, while SF in highly inclined mergers is generally more subtle (Mihos & Hernquist 1996). Blumenthal & Barnes (2018) also suggested that the impact of orbital geometry is contingent on the galaxy's gas disk size. In prograde encounters, gas inflow resulting from interactions is most pronounced in small disks entering on expansive orbits, whereas retrograde interactions favor larger gas disks. While it might be scarce for two galaxies to possess perfectly parallel orbital angular momentum in the local Universe, a prograde orbit still exerts the strongest tidal forces, inducing significant central gas inflow and consequent SFR enhancement (Zheng et al. 2020).

However, there is a scarcity of studies directly addressing the effects of SSA. Only a handful of studies exist on this topic (e.g., Mesa et al. 2014; Moreno et al. 2015). For example, utilizing the adaptive mesh refinement code RAMSES, Perret et al. (2014) conducted a comparative analysis of the SF histories of 20 merging galaxies against three isolated galaxies. They showed a more pronounced enhancement in the SFR when the spin axes of both primary and secondary galaxies were either at 90° or parallel (as opposed to having a perpendicular configuration). This trend is particularly evident in gas-rich mergers, in line with our results. Using a comprehensive set of 75 simulations of major mergers, Moreno et al. (2015) highlighted the significance of the SSA of interacting galaxies concerning interaction-induced SF enhancement. They categorized interacting galaxies based on the position angle and inclination of each galaxy's orbit into three distinct groups: (a) strongly aligned disks, (b) nearly perpendicular disks, and (c) nearly anti-aligned disks. Their findings suggest that the SSA between two interacting galaxies can be a pivotal factor in driving interaction-induced SF enhancement. Notably, primary galaxies in well-aligned interacting systems exhibit the most pronounced increase in SFE.

In the observational context, studying SSA poses more considerable challenges, primarily stemming from the inherent difficulties in accurately determining galactic spin alignments based solely on the projected disk geometry in captured images. To mitigate this challenge, Mesa et al. (2014) employed an innovative strategy. By visually inspecting the SDSS DR7 data set, they categorized approximately 2000 galaxy pairs identified by observable tidal features into "corotating" and "counterrotating" systems. This classification is contingent on whether the winding directions of the interacting galaxies are congruent or opposing. Interestingly, they discovered that corotating systems are twice as prevalent as their counterrotating counterparts. They concluded that this difference might be due to the more rapid evolutionary trajectories of the counterrotating systems. Contrary to the theoretical findings by Mesa et al. (2014), Moreno et al. (2015) demonstrated that a counterrotating sample of non-AGN galaxies show a modestly increased SFR and a larger fraction of younger stellar populations compared to corotating samples. This difference is most pronounced in closely situated pairs ($\Delta R < 12 \; h^{-1}$ kpc). It is worth noting, however, that their classification predominantly leaned on the orientation of the spiral arms, without considering the impact of disk inclination. Additionally, by bifurcating the galaxy pairs into two distinct categories—corotating and counterrotating—it becomes challenging to methodically assess how SSA influences interaction-induced SF as a continuous function.

In this study, we evaluate SSA more directly, utilizing methodology from Lee & Pen (2002) and Lee (2011). With this advancement, we investigate the dependence of interaction-induced SF strength on a combination of projected separation distance and relative spin orientation of paired galaxies. Consequently, we are the first to suggest the possibility of a systematic impact of spin configuration on the SF activity of paired galaxies through observations.

## 5. Summary and Conclusions

We explore the influence of hydrodynamic effects on interaction-induced SF as a function of projected separation and relative spin orientation between paired galaxies in the local Universe, using data from SDSS DR7 ($0.02 < z < 0.06$). Utilizing a pairing identification scheme that incorporates galaxy morphology data from Galaxy Zoo 2, we have identified 2783 S+S pairs, wherein both interacting galaxies are categorized as spiral galaxies. By employing a novel method to observationally measure the SSA (Lee & Pen 2002; Lee 2011), we parameterize the misalignment angle, $\cos \xi$, based on the inclination and position angles of galaxy disks. For the first time, we systematically present the correlation between $\cos \xi$ and interaction-induced SF.

Our main results are summarized as follows.

1. We reexamine our prior findings (Paper I) and emphasize that the presence of nearby galaxies with relatively higher SF markedly enhances the sSFR, underscoring the hydrodynamic effects on interaction-induced SF in paired galaxies.
2. The degree of spin alignment between paired galaxies represents another influential factor for interaction-induced SF. While its impact is subtler compared to the physical proximity of the pair members, it is nonetheless discernible. The interaction-induced SF is progressively augmented as configurations shift from perpendicular alignments to configurations that are more coherently aligned.
3. The correlation between spin alignment and interaction-induced SF is more pronounced when neighboring galaxies exhibit relatively higher SFRs. Similar to the correlation between projected separation and interaction-induced SF, this result suggests a hydrodynamical influence on the effect of spin alignment.
4. While one might expect that spin alignment could be caused by the proximity of pair members, the degree of spin alignment depends on neither short-range projected separation nor distance to nearby large-scale structures.

Our findings suggest that the relative spin orientation between paired galaxies is responsible for the observed variability in interaction-induced SF. While proximate galaxies with relatively higher SFRs enhance sSFR, the interaction-induced SF also displays an increased efficiency in well-aligned configurations. Only some theoretical studies have anticipated or explored this phenomenon, and until now, observational evidence has been lacking. Our study presents the first observational support for the proposition that the relative spin orientation can modify the hydrodynamic effects on interaction-induced SF. Additionally, while one might hypothesize that a well-aligned configuration is a consequence of interactions (and thus, the observed correlation between spin alignment and interaction-induced SF merely reflects simultaneous outcomes of the same interaction), we show that spin alignment persists irrespective of the short-range projected separation and the proximity to the nearest large-scale filaments. Consequently, we infer that the augmented interaction-induced SF observed in well-aligned systems is genuine. Our findings can be ascribed to two main mechanisms: (a) SF activity driven by hydrodynamic friction, wherein well-aligned interacting systems encounter more effective ram pressure, and (b) a higher probability of interactions in prograde orbits, which are generally considered conducive to enhanced SF in well-aligned configurations.

We wish to emphasize that our study is the first to suggest the necessity of investigating the effect of the SSA. The direct impact of the SSA on interaction-induced SF enhancement in paired systems has often been neglected in most previous observational and simulation studies. Thus, a comprehensive understanding of the precise physical mechanisms driving the observed heightened SF in well-aligned systems remains elusive. To bridge this gap, we intend to explore whether the SSA dependency exists in galaxies within cosmological simulations, such as IllustrisTNG, in our forthcoming series of studies.

## Acknowledgments

S.-J.Y. acknowledges support from the Mid-career Researcher Program (No. 2019R1A2C3006242) and the Basic

Science Research Program (No. 2022R1A6A1A03053472) through the National Research Foundation (NRF) of Korea. S.P. acknowledges support from the Mid-career Researcher Program (No. RS-2023-00208957) through the NRF of Korea. Funding for the SDSS has been provided by the Alfred P. Sloan Foundation, the Participating Institutions, the National Aeronautics and Space Administration, the National Science Foundation, the U.S. Department of Energy, the Japanese Monbukagakusho, and the Max Planck Society. The SDSS website is located at http://www.sdss.org/. The SDSS is managed by the Astrophysical Research Consortium for the Participating Institutions. The Participating Institutions are the University of Chicago, Fermilab, the Institute for Advanced Study, the Japan Participation Group, the Johns Hopkins University, Los Alamos National Laboratory, the Max-Planck-Institute for Astronomy (MPIA), the Max-Planck-Institute for Astrophysics (MPA), New Mexico State University, University of Pittsburgh, Princeton University, the United States Naval Observatory, and the University of Washington. The data in this paper are the result of the efforts of the Galaxy Zoo 2 volunteers, without whom none of this work would be possible. Their efforts are individually acknowledged at http://authors.galaxyzoo.org. We would like to extend our gratitude to one of our confidants, the exceptionally talented graphic designer Dyn Yu (www.instagram.com/studiodyn), for the splendid representation in Figure 2.

## ORCID iDs

Woong-Bae G. Zee https://orcid.org/0000-0003-0960-687X
Jun-Sung Moon https://orcid.org/0000-0001-7075-4156
Sanjaya Paudel https://orcid.org/0000-0002-8040-6902
Suk-Jin Yoon https://orcid.org/0000-0002-1842-4325